# The influence of Coriolis force on sedimentation of the Yellow River

Kejing Liu[1], Dawei Liu[2]

[1]Navigation College of Jimei University. Xiamen, 361021, China;
[2]Normal School, Lanzhou University of Arts and Science, Lanzhou 730000, China

*Correspondence to*: Kejing Liu (liu_kj313@126.com)

**Abstract.** In the northern hemisphere, river subjects the right bank to the pressure generated by the Coriolis force, which will increase the erosion of the river on the right bank. On the other hand, the Coriolis force also causes the sediments in the water to move to the right bank, which will increase the sediment deposition on the right bank of the river. Therefore, for rivers with low sediment content, Coriolis force will increase the erosion of river water on the right bank; for rivers with high sediment content, Coriolis force will increase the sedimentation of sediment on the right bank. It is noted that the Lanzhou section of the Yellow River has siltation of sands and pebbles to the right (south) bank. It is believed that this is caused by the Coriolis force moving the sands and pebbles to the right bank.

Due to the rotation of the Earth, rivers are subject to the Coriolis force. The prevailing view is that in the Northern Hemisphere, due to the influence of the Coriolis force, the right bank of the river is washed more strongly than the left bank, so that the right bank of the river is steep and the left bank is relatively gentle; The Southern Hemisphere is the opposite. This is Baer's law (Kleppner and Kolenkow, 1973; Liang, 2010), Baer's law is derived without taking into account the content of river sediments, implying that the water is free of sediment, so Baer's law only attaches observations to clearer rivers.

If the water body of the river is turbid and the sediment content is very large, the river not only washes away the riverbed, but also deposits a large amount of sediment into the riverbed. In this case, whether Baer's law is still true, and whether rivers with large sediment levels in the northern hemisphere are still steep on the right bank and gentle on the left bank, for example, for the Yellow River, whether the right bank is still steeper than the left bank is obviously a question worth considering. We believe that the Coriolis force not only affects the erosion of the riverbed by the river but also affects the movement and sedimentation of sediments to the riverbed, and for rivers or river sections with large sediments content, when analyzing the effect of the Coriolis force on the steepness of the riverbed, it is necessary to consider not only the effect of the Coriolis force on the erosion of the river bank by the water flow, but also the effect of the Coriolis force on the sediment deposition to the river bank. Let's discuss this issue in detail.

**1 Investigation and analysis of the Lanzhou section of the Yellow River**

Lanzhou on the Loess Plateau, between an east-west canyon, the Yellow River runs through the city from west to east. Lanzhou is the only city in the country where the Yellow River flows through the urban area. Due to the continuous movement and sediment of the river to the south bank, the riverbed continues to move north, forming a well-developed and wide multi-step terraces on the south bank, while the north bank is relatively steep, and the land is gradually narrowed. Due to the uplift movement of the land and the undercutting of the river on the soft soil, in the Lanzhou section, the riverbed of the Yellow River is very low, and the first terrace is 3-8 meters above the river surface, and the second terrace is 10-40 meters higher than the riverbed (The Figure 1 shows the terraces distribution in the Yellow River Valley, the left bank is old and sharp while the right bank is young and flat. He, Hu and Pan, 2020). In the Lanzhou section, the sediment of the Yellow River has been moving to the right bank and depositing, and the reasons for this and whether the movement and sediment of the Yellow River sediment to the right bank are general, are questions worth studying (Zhang and Zhang, 1987; Liu, Yuan and Ge, 2007; Zhang et al., 2015).

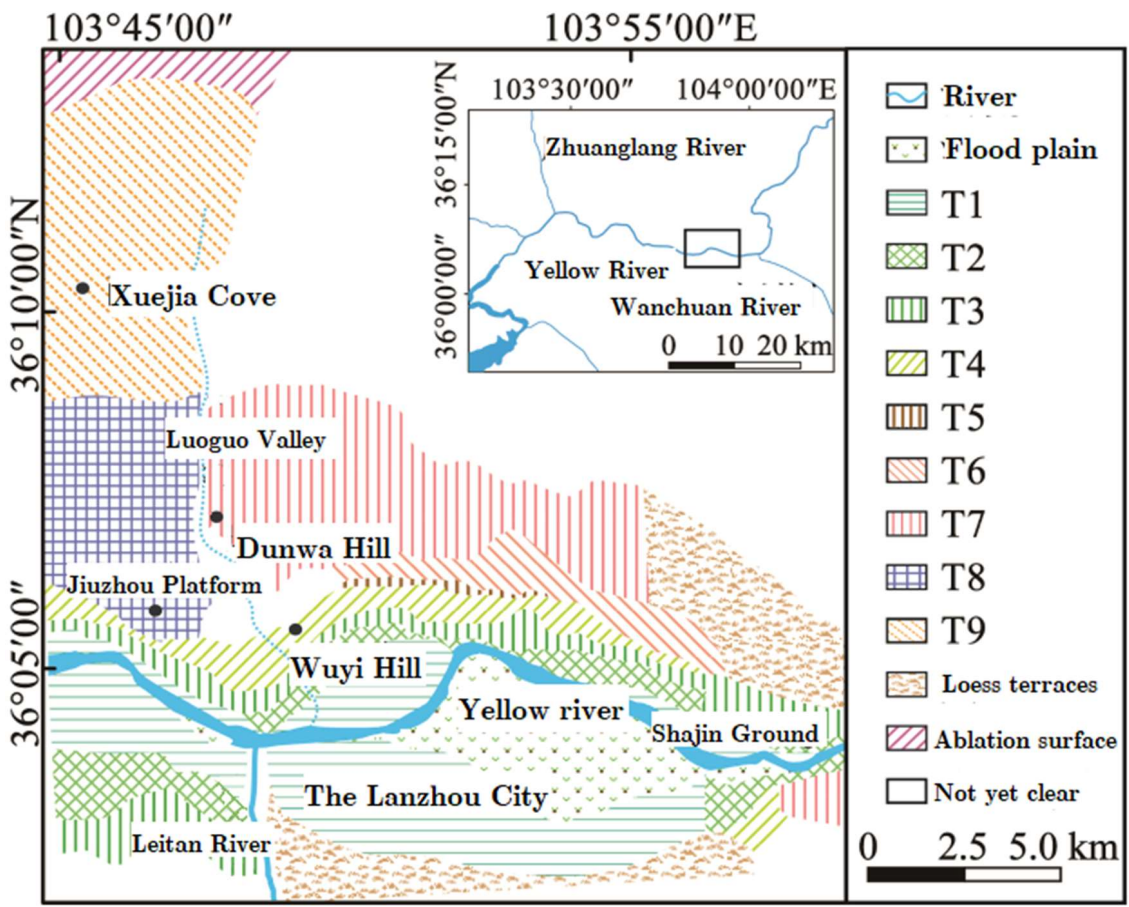

**Figure.1: The terraces distribution in the Yellow River Valley (He, Hu and Pan, 2020)**

We know that when considering the movement of air and water flow on a large scale, the Earth should be considered to be a rotating reference frame. For a river flowing from west to east on a line of latitude in the Northern Hemisphere, there is a Coriolis force pointing to the south bank (right bank) on this stream, and on Earth this inertial force is balanced by the pressure from the south bank of the river. According to Newton's third law, the river also produces a corresponding pressure on the south bank, so the south bank of the river is under greater pressure from the river, and the erosion force on the river bank increases with the increase of the water pressure received, so the south bank of the river is more seriously washed by the river (Kleppner and Kolenkow, 1973; Liang, 2010). The above discussion of Baer's law is derived without taking into account the sediment carried by river (Kleppner and Kolenkow, 1973; Liang, 2010), and does not deal with the content of sediment in river and their effects. In view of the investigation of the state of the bank in the Lanzhou section of the Yellow River, we believe that for rivers with large sediment content, the impact of sediment deposition on the riverbed cannot be ignored. To that end, we are still discussing a river flowing from west to east, taking into account the high sediment content. The inertial force of the water flow is directed to the south bank (right bank), because the density of sediment is greater than that of water, the inertial force of the Coriolis plays a role in "inertia separation" of the sediment in the water flow, the sediment in the water flow moves south under the action of the inertia force, the sediment rushes to the south bank and is deposited on the south bank, in this case, the amount of silt on the south bank can exceed the amount of erosion, so the south bank is gentle and continuously pushes towards the river center. In addition, in the cross-section of the river perpendicular to the direction of the river flow, there is also a water flow cycle that exchanges the water bodies on both sides of the riverbed - secondary flow: the water body with a large sediment content moves to the south bank and gradually sinks, and the water body with a small sediment content moves to the north bank and gradually floats up. The secondary flow of this body of water constitutes a pattern of sediment movement towards the south bank, sedimentation, and erosion of the north bank by the river (Wu, 2008; Zhan, et al., 2016). Since the sediment moves to the south bank and deposits, and the riverbed must maintain a sufficient width to maintain the flow of water, this makes the erosion effect of the river water on the north bank greater than the sediment deposition, resulting in the movement of the riverbed to the north bank. The soil of Lanzhou belongs to the loose loess soil of the Loess Plateau, which is particularly susceptible to erosion under the erosion of water currents, which aggravates the encroachment of the river on the land on the north bank, resulting in a steep north bank (Figure 2, Figure 3). In the Lanzhou section of the Yellow River, due to the continuous northward shift of the river bank, the river bank gradually approached the north bank mountain range, resulting in a very narrow land on the north bank. Therefore, the results of the investigation of Lanzhou's landform reported in the relevant literature are that "the lateral erosion of the north bank of the Yellow River is strong, and the I.-III. terraces are widely developed on the south bank of the Yellow River, forming a wide and flat terrace, of which the III. terraces are the most developed, indicating that the accumulation of the south bank of the Yellow River is strong (Liu, Yuan and Ge, 2007)." Since the Yellow River in the Lanzhou section flows along the mountains with east-west direction of on both sides, there are more straight sections of the river along the north and south directions, and the straight section of the river continues to invade the north, there is no reason to explain the phenomenon of erosion dominated the left bank and sediment dominated the right bank with bending of the river (Liang, 2010). Liu et al. (2007) believes that "the height of the same terrace on the south bank of the Yellow River is mostly higher than that of the north bank of the Yellow River, that reflecting the characteristic of the basin's high south and low north, and indicating that the accumulation of the south bank of the Yellow River is stronger than that of the north bank and the erosion or lateral erosion of the north bank is stronger than that of the south bank; At the same time, it may also indicate that the force source of tectonic deformation in Lanzhou Basin mainly comes from the southwest side and gradually passes to the north-east direction, resulting in a greater uplift of the south bank than the north bank. " The different consideration we made is that the tectonic deformation of Lanzhou and its neighbors is not easy to detect in a short period of time, so the phenomenon of "north invasion and south siltation" in the Lanzhou section of the Yellow River caused by it should also be slow and inconspicuous, while in fact, the phenomenon of "north invasion and south siltation" in the Lanzhou section of the Yellow River is obvious. Relevant literature (Zhao, He and Cheng, 2014) indicates that "between 1967 and 1998, the Lanzhou station section was in a state of erosion on the left bank and siltation on the right bank," and from the middle to the end of the 19th century, the large floodplains that bordered the south bank of the Yellow River were rapidly covered by buildings. Therefore, we believe that the Coriolis inertial force produced by the rotation of the earth and the flow of the river has the effect of "inertial separation" for rivers with large sediment content (most typically the Yellow River) (Kleppner and Kolenkow, 1973; Liang, 2010), which plays an important role in influencing the characteristics of the Yellow River bed in the Lanzhou section and constitutes an important factor determining the landform of Lanzhou today. The "northward invasion and south siltation" of the Yellow River bed in the Lanzhou section is not an isolated case, and the relevant literature (Zhou, et al., 2020) indicates that there is also a left bank erosion in the Ningxia section of the Yellow River, and the phenomenon of siltation in other locations. Ji, Fan and Zhai et al (2020) found that between 1988-2018, the section from SanHe Hukou to Zhaojun Grave in the Inner Mongolia of the Yellow River ($R_3$ of the Fig.4) oscillated to its left bank totally for it is the section with most sediments input, the Fig 5 show range of wiggle of every section of the Yellow River flows through Ningxia and Inner Mongolia and the Table 1 show the oscillating distance to the left bank and the right bank of them. The river banks became more stable from 2013 for the sediments reduced.

According to the investigation of multiple sections of the Yellow River (Wells, 2009), during the normal water level of the Yellow River, especially during the dry season, a large floodplain will be exposed on the side adjacent to the right bank, the mainstream is far from the right bank, and the floodplain is submerged during the flood period, and the river fills the riverbed. We believe that this phenomenon should be a sign of the "northward invasion and southern siltation" of the Yellow Riverbed. This phenomenon is evident in the Lanzhou section of the Yellow River.

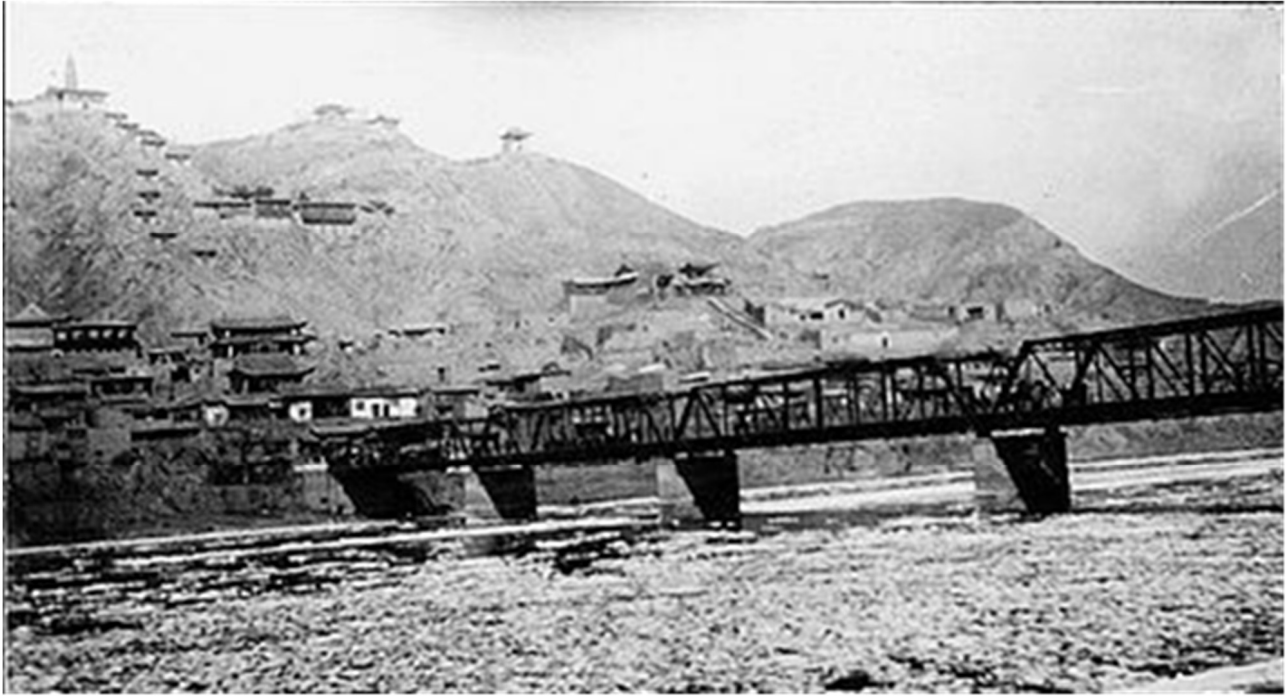

**Figure.2: Steep north side and gentle inclined south side with shingles of the Yellow River (an old photo of the Lanzhou Zhongshan Bridge)**

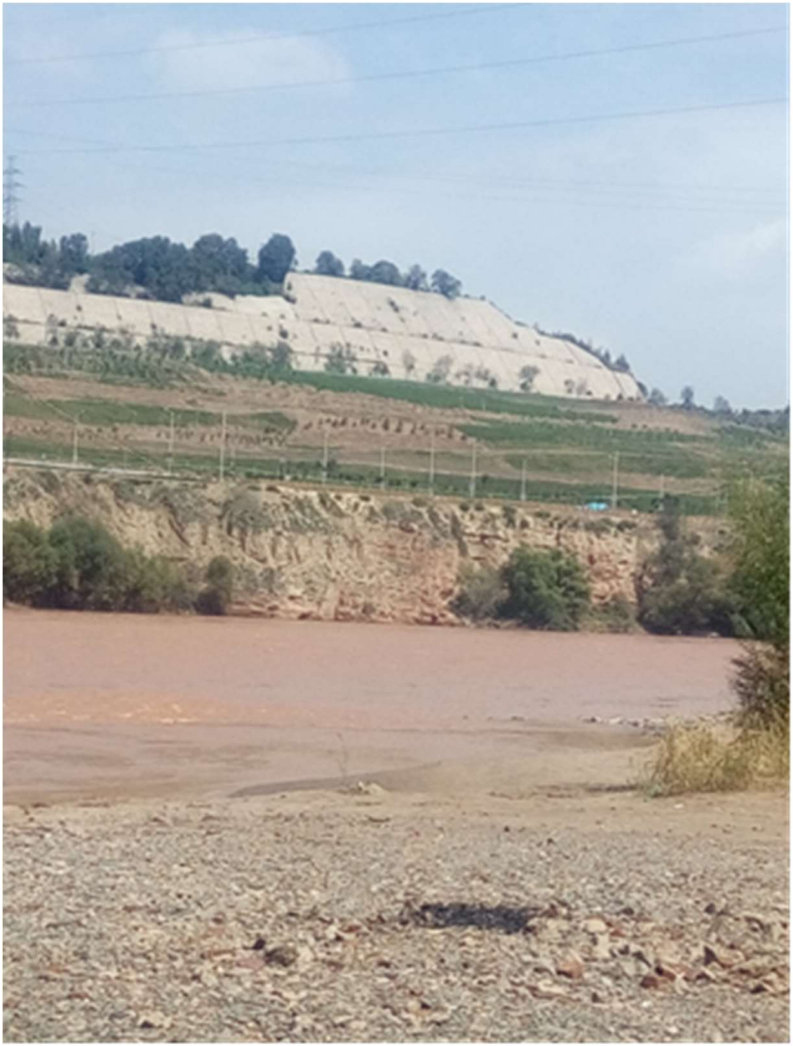

**Figure.3: Scouring in the north side, and depositing in the south side (took on a section of the Yellow River passed Xingang city of Lanzhou)**

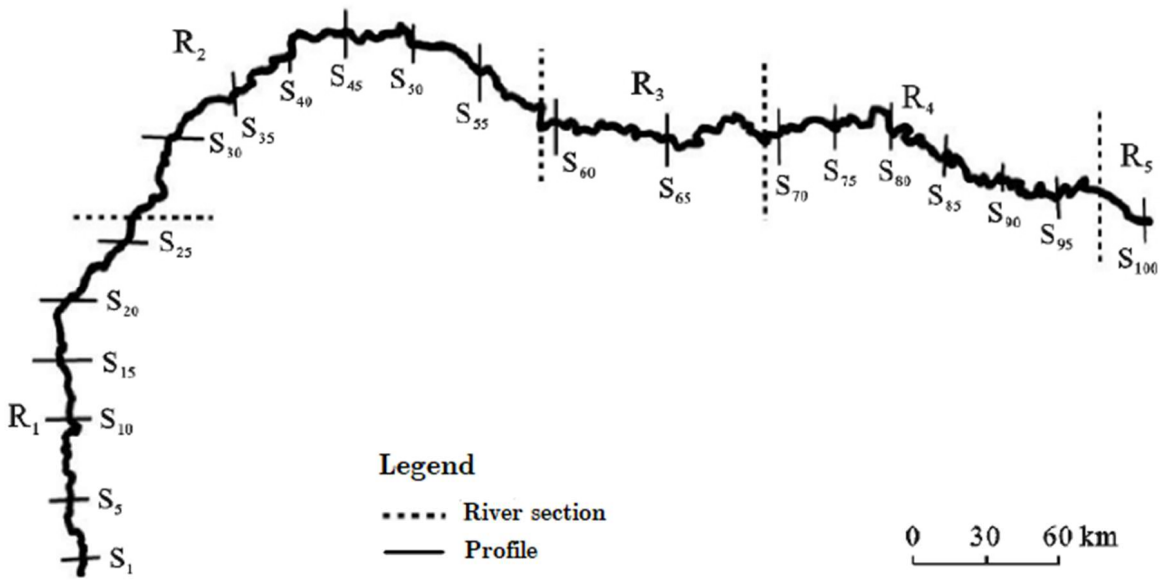

**Figure.4:** The four ($R_1$, $R_2$, $R_3$ and $R_4$) sections of the Yellow River of Ningxia and Inner Mongolia (Ji, Fan and Zhai et al, 2020)

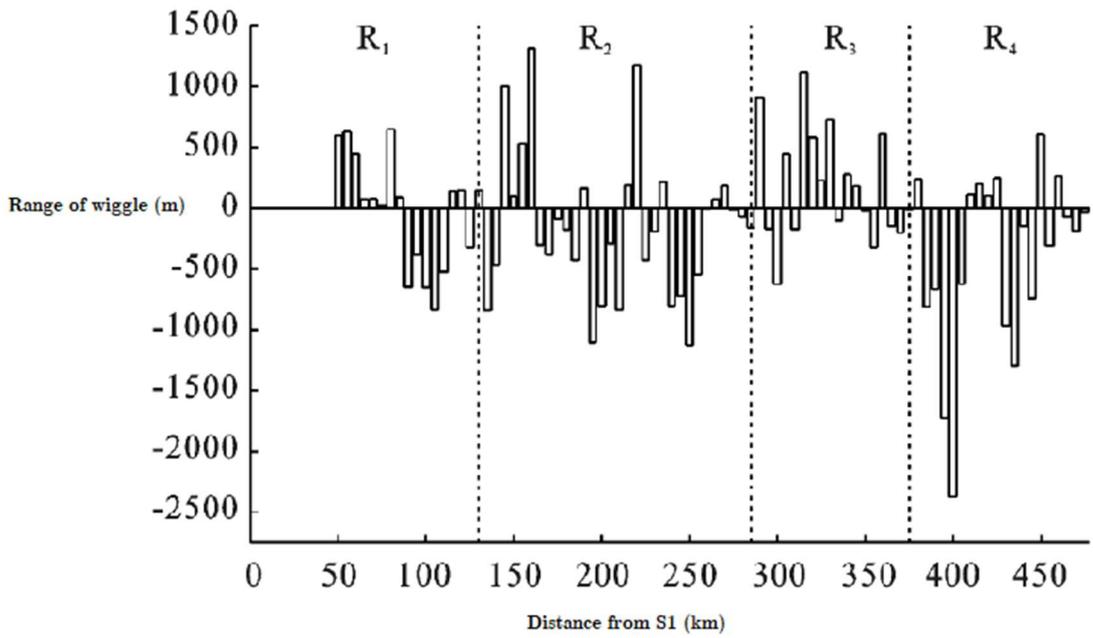

**Figure 5:** Range of wiggle of every profile in every section (Ji, Fan and Zhai et al, 2020)

**Table 1:** The oscillating distance to the left bank and the right bank of every section (Ji, Fan and Zhai et al, 2020)

| Section | R1 | R2 | R3 | R4 |
|---|---|---|---|---|
| Oscillating to the left bank | 36.19 | 209.94 | 92.92 | 39.60 |
| Oscillating to the right bank | 39.36 | 270.85 | 79.43 | 73.04 |

## 2 Mathematical model of sediment deposition to the south bank of the Yellow River

Under the action of the Coriolis force, a sediment particle with a mass of *m* moves at an acceleration *a* toward the south bank. If the Coriolis force it receives is $F_c$, and the resistance it receives is $F_r$, then there is $ma = F_c - F_r$, and if we think of the sediment particle as a sphere with a density of $\rho$ and radius of *R*, then there can be $\frac{4}{3}\pi R^3 \rho a = a_c \frac{4}{3}\pi R^3 \rho - (\frac{4}{3}\mu)\pi R^2$, in where $\mu$ is the scale factor, simplified to obtain

$$a = a_c - \frac{\mu}{R\rho} \quad (1)$$

In Equation (1), taking into account the size of the Coriolis acceleration $a_c = 2\omega v_r \sin\alpha$ we write it as $a_c = Lv_r$ ($L = 2\omega \sin\alpha$), where $\omega$ is the angular velocity of the Earth's rotation, $v_r$ is the velocity of the Yellow River, and $\alpha$ is the angle between the river velocity and the north-south direction of the Earth's axis of rotation, so the formula (1) becomes

$$a = \frac{dv}{dt} = Lv_r - \mu \frac{1}{R\rho} \quad (2)$$

By formula (2) can be obtained

$$v = v_0 + (Lv_r - \mu \frac{1}{R\rho})t \quad (3)$$

From $v_0 = 0$, equation (3) becomes

$$v = (Lv_r - \mu \frac{1}{R\rho})t \quad (4)$$

Both equation (1) and equation (4) show us that under the action of the Coriolis force, the greater the flow rate of the water flow, the easier it is to deposit sediment and pebbles on the riverbank; The larger the scale and density, the more likely it is that sediment particles will be deposited on the riverbank. Since in the Northern Hemisphere, the Coriolis force is directed to the right bank of the river, the sediment and pebbles in the river are moved to the right bank of the river (Figure 2, Figure 3). Due to the large sediment content of the Yellow River and the secondary flow with higher speed on the surface of the river than on the bottom (Einstein, 1926), the amount of sediment moving from the river to the right bank is correspondingly large, which can make the sediment deposition of the river on the right bank exceed the amount of sediment washed by the river on the right bank. In the Lanzhou section of the upper reaches of the Yellow River, the river flow rate is large, and when the Yellow River enters the Lanzhou section, the sediment content in the water is already very high, so in the Lanzhou section of the Yellow River, the phenomenon of sediment deposition in the river to the south bank is very obvious. Equation (2) is a differential equation that reflects the Coriolis force that separates sediment particles from flowing water and transfers them to the riverbank.

**3 Ideal experiments simulating sediment deposition of sediment to the south bank of the Yellow River using sediment settlement in still water**

Using the exact same derivation process from equations (1) to (4), we can obtain a formula for the settlement of particulate matter in a stationary fluid (e.g., in water in a graduated cylinder) under the action of gravity.

The acceleration of settlement is

$$a = g - \mu \frac{1}{R\rho} \qquad (5)$$

When the finishing speed is not reached, the speed of settlement is

$$v = v_0 + (g - \mu \frac{1}{R\rho})t \qquad (6)$$

In Formula (6), when the particulate matter is attached to the surface of the still water, it can be taken $v_0 = 0$, so there is

$$v = (g - \mu \frac{1}{R\rho})t \qquad (7)$$

The two equations, (5), (7) are easy to verify, comparing formula (5) with formula (1) or (2), formula (7) with formula (4), it can be seen that under the action of gravity, the sedimentation ideal experiment of particles with a density of $\rho$ and radius of $R$ in the water in the measuring cylinder can simulate the effect of the Coriolis force on the sedimentation of sand and gravel in the river water (Wells, 2009). By the way, (5), (6), (7) the three equations can also be explained that: in the atmosphere, fine dust falls very slowly, so in the landing process it is easy to be carried far away or brought high again by the various air currents that often appear in the atmosphere. This means that the finer the dust, the easier it is to spread and the less likely it is to land. This fact is a verification of the correctness of the (5), (6), (7) forms, and thus of the correctness of the equations (1) to (4). Literature (Zhao, He and Cheng, 2014) shows that "from 1967 to 1998, the section of Lanzhou Station was in a state of erosion on the left bank and siltation on the right bank, and with the joint dispatch of Longyangxia and Liujiaxia reservoirs in 1986, the flow distribution tended to be uniform during the year, the amount of suspended mass sand transport decreased, the particle size became thinner, and the siltation of the Lanzhou section of the Yellow River gradually became stable." This is consistent with Equations (1) to (4), from which it can be seen that a decrease in the peak flow rate and a smaller radius of the suspended matter particles reduce the effect of the Coriolis force in shifting the suspended particles to the south bank.

Formula (1) is derived by considering the Earth as a non-inertial frame, using Newton's second law in form in the non-inertial frame (Zhan, Cheng and Liu, 2016). Equation (5) is the law of motion of particulate matter in the still water contained in the graduated cylinder, and in deriving this law, the Earth can be used as an inertial frame, and Equation (5) is derived from Newton's second law, which is true only in the inertial frame. Equations (1) and (5) are formally identical, which is a manifestation of the equivalence of the inertial force of an object with acceleration and the gravitational force of a gravitational field on the object in a local range (Li, 2007; Zhu, 2000). Using the similarity between the law of sediment settlement in still water under gravity and the law of sediment deposition to the riverbank under the action of the Coriolis force, we can make a simple and rough study of the sediment to the river bank under the action of the Coriolis force through the experiment of sediment settlement in still water.

## 4 Conclusion

Observations, theories and ideal experiments can show that in the northern hemisphere, the Inertial Force of Coriolis is the reason why the scouring force on the right bank of the river is greater than that of the left bank, and the Inertial Force of Coriolis is also the reason why the sediment in the river is easy to deposit on the right bank, and the two opposite effects of the inertial force on the river bank which play a leading role, lies in the sediment content of the river, the former role has nothing to do with the sediment content, the latter role increases with the increase of sediment content, in the case of less sediment content, the former's role exceeds the latter, which is the situation described by Baer's law. In the case of a large sediment content, the effect of the latter can exceed that of the former, which is due to the separation effect of the inertia force on the sediment. The summer of 2022 is very hot and with serious drought in large area of the northern hemisphere, and many rivers withered to very low water level, it is an unlucky climatic extreme event but also a chance for observing the form of the banks and distribution of sands and pebbles.